\documentclass[twocolumn,showpacs,amssymb,nobibnotes, aps,prl]{revtex4-1}

\usepackage{enumerate,amsmath}
\usepackage{graphicx}
\begin{document}

\title{Robust and Scalable Scheme to Generate Large-Scale Entanglement Webs}%

\author{Keisuke Fujii}
\author{Haruki Maeda}
\author{Katsuji Yamamoto}
\affiliation{
Department of Nuclear Engineering, Kyoto University, Kyoto 606-8501, Japan}

\date{\today}
\begin{abstract}
We propose a robust and scalable scheme to generate an $N$-qubit $W$ state
among separated quantum nodes (cavity-QED systems) by using linear optics and postselections.
The present scheme inherits the robustness of the Barrett-Kok scheme 
[Phys. Rev. A {\bf 71}, 060310(R) (2005)].
The scalability is also ensured in the sense that
an arbitrarily large $N$-qubit $W$ state can be generated
with a quasi-polynomial overhead $\sim 2^{O[(\log _2 N)^2]}$.
The process to breed the $W$ states, which we introduce to achieve the scalability,
is quite simple and efficient, 
and can be applied for other physical systems.
\end{abstract}


\maketitle

{\it Introduction.---}
So far tremendous efforts have been paid for
experimental realizations of quantum information processing (QIP), 
and, for example, control of a few qubits has been performed
in cavity QED, ion traps, etc.
It, however, seems difficult to increase
the number of qubits dramatically within a single physical system.
In order to realize large-scale QIP,
we have to develop a novel way to integrate individual physical systems scalably.
Furthermore, for communication purposes,
quantum information has to be shared among separated quantum nodes.
To meet these requirements, {\it distributed} QIP,
where stationary qubits are entangled by using flying qubits (photons), 
seems to be very promising \cite{Cabrillo99,Bose99,Benjamin09,Duan10}.
A lot of protocols have been proposed so far
for remote entangling operations 
and probabilistic two-qubit gates \cite{REO,Barrett05,LD0506}.
The Barrett-Kok scheme is particularly promising, since 
it is fully scalable and robust against experimental imperfections \cite{Barrett05}. 
Experiments of the remote entangling operations (or probabilistic two-qubit gates)
between separated qubits have also been done 
in both atomic ensembles \cite{Chou05} and trapped single atoms \cite{MOM0709}.
They are important ingredients for fault-tolerant distributed quantum computation
\cite{Barrett05,Duan05,Li10,FT10}.

Multipartite entanglement is not only a key ingredient 
for quantum communication, but also an important clue
to understand nature of quantum physics.
There are a lot of classes of multipartite entanglement, for example,
GHZ (Greenberger-Horne-Zeilinger) states \cite{GHZ},
cluster states \cite{Cluster}, and $W$ states \cite{Wstate},
which cannot be transformed each other 
under local operations and classical communication (LOCC).
Among them, the $W$ states,
\begin{eqnarray*}
| W_{N}\rangle =\frac{1}{\sqrt{N}} ( | 1 0 0 \cdots 0 \rangle + |010 \cdots 0\rangle
\cdots +|000 \cdots 1 \rangle, 
\end{eqnarray*}
are quite robust
in the sense that any pairs of qubits are still entangled, 
even if the rest of the qubits are discarded \cite{WstateRobust}.
This web-like property is very fascinating
as universal resources, i.e., {\it entanglement webs}, for quantum communication.
There are several protocols, which use the $W$ states
for quantum key distribution, teleportation, leader election, and information splitting \cite{WstateApp}.
Furthermore, inevitable decoherence in sharing the $W$ states
can be counteracted by using a novel scheme of purification \cite{Miyake05}.
The preparation of the $W$ states by using optics has been discussed so far extensively
both theoretically and experimentally \cite{WstateOpt}.
It has been also discussed in other systems,
such as cavity QED and ion traps \cite{ExOther}.
Nevertheless none of them seems to be fully scalable.
That is, overhead required for sharing an $N$-qubit
$W$ state scales exponentially in the number of qubits $N$, or
the $W$ state is prepared in a single system,
which cannot be used for quantum communication among separated quantum nodes.

In this paper, we develop a {\it robust and scalable} scheme to
generate the $N$-qubit $W$ state
by using separated cavity-QED systems and linear optics.
The present scheme is scalable
in the sense that an arbitrarily large $N$-qubit $W$ state can be generated 
among separated quantum nodes with only 
a quasi-polynomial overhead $\sim 2^{O[(\log _{2} N)^2]}$.
In the following, 
we first develop an efficient way to generate the four-qubit $W$ state $|W_4\rangle$
by following the concept of the Barrett-Kok scheme \cite{Barrett05},
which is quite robust against the experimental imperfections.
The success probability to obtain the $|W_{4}\rangle$
is significantly high to be 1/2.
Then, by using the four-qubit $W$ states as {\it seeds},
we can {\it breed} an arbitrarily large $W$ state in an economical way,
where the two $|W_{N}\rangle$'s are converted
to one $|W_{2(N-1)}\rangle$ probabilistically by accessing only two qubits.
In contrast to classical webs, 
where a local connection does not result in a global web structure,
this property of entanglement webs is a genuine quantum phenomenon.
Even if the conversion fails,
the two $|W_{N-1}\rangle$'s are left and can be recycled.
This breeding method is quite simple and economical,
and can be applied to other physical systems, such as polarization qubits in optics \cite{WstateOpt}. 

\begin{figure}
\centering
\includegraphics[width=70mm]{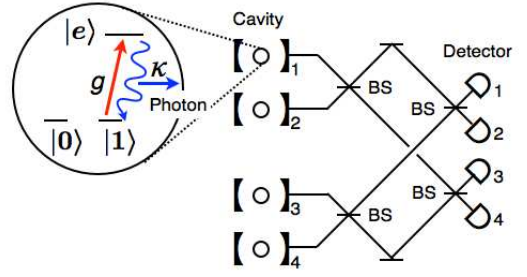}
\caption{Three-level atoms are embedded in cavities. The two long-lived states $|0\rangle$ 
and $|1\rangle$ are used as a qubit, and the transition between the states $|1\rangle  \leftrightarrow 
|e\rangle$ is coupled to the cavity mode. The output modes are mixed with 
four 50:50 BS's, and measured by photodetectors.}
\label{Setup}
\end{figure}

{\it Four-qubit $W$ state (seeding).---}
We consider four three-level atoms, each of which 
is embedded in a separated cavity. 
The two long-lived states of the atom  $|0\rangle$ and $|1\rangle$
are used as a qubit, where
only the state $|1\rangle$ is coupled to the excited state $|e\rangle$,
whose transition frequency is equal to that of the cavity mode (see Fig. 1).
The output fields of the cavities 
are mixed with 50:50 beam splitters (BS's) and measured by photodetectors.
The effective Hamiltonian of the system is given by
\begin{eqnarray*}
H = \sum _{i=1}^{4} \frac{g_{i}}{2} ( |1 \rangle _{ii} \langle e| \hat{c}_{i}^{\dag} + {\rm H.c.})
-  i \sum _{i=1}^{4} \kappa _{i} \hat{c}_{i}^{\dag}\hat{c}_{i},
\label{Hamiltonian}
\end{eqnarray*}
where $g_{i}$ denotes 
the coupling between the $|1\rangle _{i} \leftrightarrow |e\rangle _{i}$ transition
and the $i$th cavity mode $\hat{c}_{i}$.
The cavity photon leaks to the output mode 
with rate $2 \kappa _{i}$ ($\kappa _{i} > g_{i}$),
which is treated as the non-Hermitian term
by following the quantum jump approach \cite{Jump}.
For simplicity, the cavity parameters are set to $g_{i}=g$
and $\kappa _{i} =\kappa$ ($i=1,2,3,4$).
As shown in Fig. \ref{Setup}, the output modes are mixed 
by using the four 50:50 BS's.
Thus the four detector modes $\hat{a}_{i}$ are given
in terms of the cavity modes $\hat{c}_{i}$ by
\begin{eqnarray*}
\hat a_{1} &=& (\hat c_1 +\hat c_2+\hat c_3+\hat c_4 )/2, \;
\hat a_{2} = (\hat c_1 -\hat c_2+\hat c_3-\hat c_4 )/2,
\\
\hat a_{3} &=& (\hat c_1 +\hat c_2-\hat c_3-\hat c_4 )/2, \;
\hat a_{4} = (\hat c_1 -\hat c_2-\hat c_3+\hat c_4 )/2.
\end{eqnarray*}

The procedure to obtain the four-qubit $W$ state $|W_{4}\rangle$ is as follows.
We first prepare the initial state of the atoms
as $| \Psi (0)\rangle =  (| 0 \rangle + |e\rangle )^{\otimes 4}/4 $
by using $\pi$-pulses.
Then, we wait for a sufficiently long time $t_{\rm w}$ to detect photons.
Before proceeding to the second round, 
each qubit is flipped as $|0\rangle \leftrightarrow |1\rangle$,
and the state $|1\rangle$ is excited to $|e\rangle$ by a $\pi$-pulse.
Then wait again for $t_{\rm w}$ to detect photons.
If three and single detector clicks, or vice versa, 
are observed at the first and second rounds, 
respectively, 
the $|W_{4}\rangle$ is obtained up to unimportant phase factors,
which can be removed by using local operations. 

Let us see in detail how the $|W_{4}\rangle$ is generated,
and calculate the success probability.
For concreteness, we consider
the case, where the 1st, 2nd and 3rd detectors 
are clicked at $t_1,t_2,t_3$ ($t_1<t_2<t_3$),
respectively, in the first round. 
In the second round, the $4$th detector is clicked at $t_{4}$.
The state conditioned on the first three clicks 
is given up to normalization as
\begin{eqnarray*}
&& | \Psi (t_1,t_2,t_3) \rangle  
\nonumber  \\
 &=&  (2\kappa)^{3/2}\hat a_{3}  e^{ - i H (t_3 -t_2) }  \hat a_{2}  e^{ - i H (t_2 -t_1)} \hat a_{1}  e^{ - i H t_1 } |\Psi (0)\rangle
 \nonumber \\
 &=&
\frac{(2\kappa) ^{3/2}}{8}\alpha(t_3) \alpha(t_2) \alpha (t_1) 
\Bigl [
\mathcal{W} (|1,0\rangle, |0,0\rangle) 
 \nonumber \\
 &&
+
\alpha (t_3) \mathcal{W} (|1,0\rangle, |1,1\rangle) 
+
 \beta (t_3) \mathcal{W}( |1,0\rangle , |e,0\rangle)
 \Bigr],
\end{eqnarray*}
where $|a,b\rangle$ ($a \in \{0,1,e\}$ and $b \in \{0,1\}$) indicates the states of
the atom $|a\rangle$ and photon $|b\rangle$, respectively,
for the combination
$\mathcal{W}( |A\rangle , |B\rangle) \equiv (|A\rangle|A\rangle|A\rangle|B\rangle-|A\rangle|A\rangle|B\rangle|A\rangle
-|A\rangle|B\rangle|A\rangle|A\rangle+|B\rangle|A\rangle|A\rangle|A\rangle)/2$.
The coefficients $\alpha (t)$ and $\beta (t)$ are the solutions to the 
Schr{\"o}dinger equation:
\begin{eqnarray*}
\alpha (t) &=& -ig/(2\sqrt{\kappa ^2 - g^2}) ( - e^{ \omega _{+} t} + e^{ \omega _{-} t}),
\\
\beta (t) &=& g^2/(4 \sqrt{\kappa ^2 -g^2}) ( - e^{ \omega _{+} t}/\omega _{+}  + e^{ \omega _{-} t} /\omega _{-}),
\end{eqnarray*}
where $\omega _{\pm} = (- \kappa \pm \sqrt{\kappa ^2 -g^2})/2$.
The probability of such an event is given by
\begin{eqnarray*}
p(t_1,t_2,t_3) 
&=& | \langle \Psi (t_1,t_2,t_3)|\Psi (t_1,t_2,t_3)\rangle |^2.
\end{eqnarray*}
For the sufficiently long $t_{\rm w}$ ($\gg 1/|\omega _{\pm}|$), the states
$|e,0\rangle$ and $|1,1\rangle $ decay 
to $|1,0\rangle$ incoherently.
The postmeasurement state at $t_{\rm w}$ is given by
\begin{eqnarray*}
\rho(t_{\rm w}) =\mathcal{N} \left[ \rho _{\mathcal{W}} (|1,0\rangle , |0,0\rangle)
+
 |\alpha (t_{\rm w})|^2  |1,0\rangle \langle 1,0|^{\otimes 4} \right],
\end{eqnarray*}
where $\rho _{\mathcal{W}}(|A\rangle , |B\rangle)= \mathcal{W} (|A\rangle , |B\rangle)  \mathcal{W} (|A\rangle , |B\rangle)^{\dag}$, and 
$\mathcal{N}=(2\kappa) ^{3}|\alpha(t_3) \alpha(t_2) \alpha (t_1)|^2/p(t_1,t_2,t_3)$.

Before proceeding to the second round,
each qubit is flipped as $|0\rangle \leftrightarrow |1\rangle$,
and the state $|1\rangle$ is excited to $|e\rangle$
similarly to the first round.
Then, the initial state of the second round
is given by
\begin{eqnarray*}
\rho'(0)  = \mathcal{N} \left[ \rho _{\mathcal{W}} (|0,0\rangle , |e,0\rangle)
+
 |\alpha (t_{\rm w})|^2  |0,0\rangle \langle 0,0|^{\otimes 4} \right].
\end{eqnarray*}
Since the first term has exactly one excitation,
by observing the single detector click at $t_4$,
the second term is removed in this round.
Finally we obtain the four-qubit $W$ state $|W_{4}\rangle$ for the atoms.
The joint probability for the first three clicks and the second single click is calculated as
\begin{eqnarray*}
p(t_1,t_2,t_3,t_4) &=&p(t_4 | t_1,t_2,t_3)p(t_1,t_2,t_3)
\nonumber \\
&=&{\rm Tr} [ 2\kappa \hat a _{4}^{\dag} \hat a _{4} e ^{ -i H t_4} \rho' (0) e^{ iHt_4}] p(t_1,t_2,t_3)
\nonumber \\
&=& (2\kappa)^4| \alpha (t_1)\alpha (t_2)\alpha (t_3)\alpha (t_4)|^2/64.
\end{eqnarray*}
For $t_{\rm w}$ sufficiently long $t_{\rm w}$ ($\gg 1/|\omega _{\pm}|$),
the success probability is calculated as
\begin{eqnarray*}
 \prod _{i=1}^{4} \int _{0} ^{t_{\rm w}} dt_i
\frac{(2\kappa)^4}{64}| \alpha (t_1)\alpha (t_2)\alpha (t_3)\alpha (t_4)|^2 = \frac{1}{64},
\end{eqnarray*}
where the sum over the orderings of $t_1,t_2,t_3$ are also taken.
The number of the events with three and single detector clicks, or vice versa, 
at the first and second round, respectively, is $2\times 4 \times 4$.
Thus the total success probability is unexpectedly high to be $p=1/2$.
This success probability can also be understood by the fact that
the initial state $(|0\rangle + |1\rangle) ^{\otimes 4}/4$
contains two types of the $W$ states (i.e., $|0001\rangle \cdots$ and $|1110\rangle \cdots$) with each probability 1/4.
Then in the present setup, we can fully extract the $W$ states
by virtue of the highly symmetric detector modes.
This method inherits the robustness of the 
Barrett-Kok scheme \cite{Barrett05};
the detector inefficiency and photon loss do not deteriorate the fidelity,
but only decrease the success probability.
The success probability scales like
$p=( \eta _{d} \eta _{l}) ^4 /2$, where $\eta _{d}$ and $1-\eta _{l}$
denote the detector efficiency and photon loss rate, respectively.
Other imperfections such as decoherence of the qubits, detector dark counts,
and mode mismatchings would not deteriorate the fidelity crucially
for a specific physical system such as the NV-diamond system,
as discussed in Ref. \cite{Barrett05}.  

The above process to prepare the $|W_{4}\rangle$ is viewed as a single concatenation of entangling operation,
$|1\rangle \rightarrow |10\rangle + |01\rangle$ and $|0\rangle \rightarrow |00\rangle$.
It can be extended straightforwardly to generate
an $N$-qubit ($N=2^L$ with an integer $L$)  $W$ state $|W_{N}\rangle$ with probability $N/2^{N-1}$
by using a similar setup.
The detector modes are  given by
$\hat a^{(L)}_{j} = A^{(L)} _{ij} \hat c_{i} / \sqrt{N}$
in terms of an $N \times N$ matrix $A^{(L)}$ 
generated recursively by
\begin{eqnarray*}
A^{(L+1)} = 
\left (\begin{array}{cc}
A^{(L)} & A^{(L)} 
\\
A^{(L)}& - A^{(L)} 
\end{array} \right),
\end{eqnarray*}
where $A^{(0)}= 1$. 
Then single and $N-1$ clicks, or vice versa, at the first and second rounds, respectively,
result in the $|W_{N}\rangle$.
More generally,
if we observe $m$ and $N-m$ clicks at the first and second rounds, respectively,
we can obtain the Dicke-symmetric state \cite{Dicke}:
\begin{eqnarray*}
|D_{m,N-m}\rangle = \sum _{i} \mathcal{S}_{i} (|0\rangle ^{\otimes m} |1\rangle ^{\otimes N-m}\rangle )/
\sqrt{{\rm C}_{N-m}^{m}},
\end{eqnarray*}
where $\{ \mathcal{S}_{i}\}$ denotes the set of all distinct combinations of the qubits,
and ${\rm C}_{N -m}^{m}= N!/[m! (N-m)!]$.
With the increasing number of qubits $N$, however,
the success probability $\sim 2^{-O(N)}$ diminishes exponentially.

\begin{figure}
\centering
\includegraphics[width=85mm]{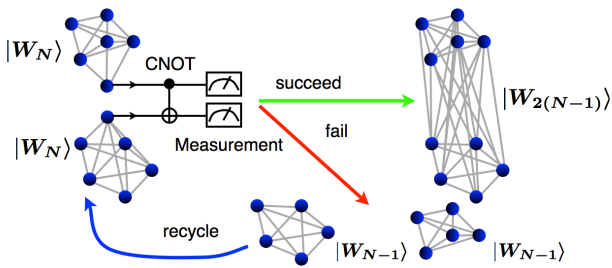}
\caption{Economical breeding. Two $|W_{N}\rangle$'s are converted 
to a $|W_{2(N-1)}\rangle$ probabilistically. Even if the conversion fails,
the two $|W_{N-1}\rangle$'s are left and can be recycled. }
\label{Breeding}
\end{figure}
{\it Economical breeding.---}
We next show that the four-qubit $W$ states
are sufficient to generate an arbitrarily large $W$ state
with a quasi-polynomial overhead, introducing 
an economical breeding (Fig. \ref{Breeding}).
Suppose that we have obtained the $N$-qubit $W$ states:
\begin{eqnarray*}
| W_{N} \rangle =\frac{1}{\sqrt{N}} 
|1\rangle ^{({\rm a})}  | 0 _{N-1} \rangle  +\sqrt{\frac{N-1}{N}} |0 \rangle ^{({\rm a})} |W_{N-1}\rangle,
\end{eqnarray*}
where the qubit labeled by (a) is used as an ancilla for the breeding, 
and $|0_{n}\rangle \equiv | 0\rangle ^{\otimes n}$.
Then, the two $N$-qubit $W$ states can be rewritten as
\begin{eqnarray*}
&&|W_{N} \rangle | W_{N} \rangle  
\\
&=&
\frac{1}{N} 
|11\rangle ^{({\rm a})}  | 0 _{2(N-1)}\rangle 
+\frac{\sqrt{N-1}}{N}|10 \rangle ^{({\rm a})} |W_{N-1}\rangle |0_{N-1}\rangle
\nonumber \\
&&
+\frac{\sqrt{N-1}}{N}|01 \rangle ^{({\rm a})} |0_{N-1} \rangle |W_{N-1}\rangle
\\
&&
 +\frac{N-1}{N}|00 \rangle ^{({\rm a})} |W_{N-1} \rangle |W_{N-1} \rangle,
\end{eqnarray*}
where the two ancilla qubits in the $W$ states 
are moved to the first two-qubit Hilbert space labeled by (a).
Here, we perform a controlled-NOT (CNOT) gate between the two ancilla qubits,
and measure the second ancilla qubit in the $Z$ basis.
If the measurement outcome is 1,
the postmeasurement state is given by
\begin{eqnarray*}
\frac{1}{\sqrt{2}}(|11\rangle ^{({\rm a})} |W_{N-1}\rangle |0_{N-1}\rangle
+
|01\rangle ^{({\rm a})} |0_{N-1} \rangle |W_{N-1}\rangle).
\end{eqnarray*}
The probability for obtaining such an outcome
is $(N-1)/N^2$.
Next, by measuring the first ancilla qubit in the $X$ 
basis and performing local operations properly depending on the outcome,
we can convert the two $N$-qubit $W$ state to the $2(N-1)$-qubit $W$ state:
\begin{eqnarray*}
\frac{1}{\sqrt{2}} (|W_{N-1}\rangle |0_{N-1}\rangle+|0_{N-1} \rangle |W_{N-1}\rangle)
= |W_{2(N-1)}\rangle.
\end{eqnarray*}
This indicates a good property of entanglement webs; 
a local connection produces a global web structure.

Alternatively, if the outcome of the first measurement for the second ancilla qubit is 0, 
we have
\begin{eqnarray*}
\frac{ 
|10\rangle ^{({\rm a})}  | 0 _{2(N-1)}\rangle 
 + (N-1)|00 \rangle ^{({\rm a})} |W_{N-1} \rangle |W_{N-1} \rangle }
 {\sqrt{N^2-2N+2}}.
\end{eqnarray*}
Then, by measuring the first ancilla qubit in the $Z$ basis with the outcome 0,
the two $|W_{N-1}\rangle$'s are left,
which can be recycled to generate the $|W_{2(N-2)} \rangle$.
The joint probability to obtain such outcomes $(0,0)$
is $(N-1)^2 /N^2$.

Notice in the above that in order to grow the size of the $W$ state,
$2(N-1) > N$ is required, i.e., $N>3$.
Thus starting from the four-qubit $W$ states,
we can breed an arbitrarily large $W$ state
by repeating the conversion process.
Not only with an even number of qubits,
we can also obtain the $W$ state with an odd number of qubits
as byproducts when the conversion fails.

In the cavity-QED setup such as Fig. \ref{Setup},
instead of the above procedure (CNOT and measurements),
the original Barrett-Kok scheme can be used to project 
the ancilla qubits to the subspace spanned 
by $\{ |10\rangle ^{({\rm a})},|01\rangle ^{({\rm a})}\}$.
Then, if the projection is successful with probability $(N-1)/N^2$,
the two $|W_{N}\rangle$'s are converted to the $|W_{2(N-1)}\rangle$.
In the failure case, then, if the ancilla qubits (atoms) 
are confirmed to be in the $|00\rangle ^{(a)}$ 
by measuring them directly,
the two $|W_{N-1}\rangle$'s are left for recycling.
Even when the detector inefficiency and photon loss are considered,
the conversion probability is diminished by only $(\eta _{d} \eta _{l}) ^2$.

\begin{figure}
\centering
\includegraphics[width=60mm]{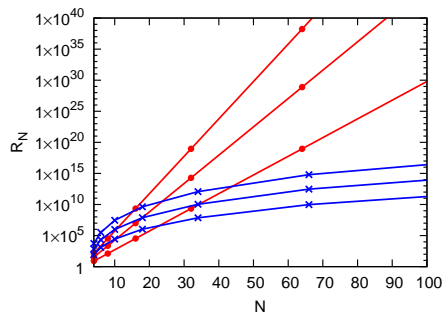}
\caption{The overheads $R_{N}$ for the concatenated entangling (red $\circ$) and the breeding (blue $\times$),
respectively, are plotted as functions of the number of qubits $N$,
where $ \eta _{d} \eta _{l} = 0.5, 0.7 ,1$ from top to bottom.}
\label{WResource}
\end{figure}

The success probability for the breeding sequence 
$|W_{4}\rangle \rightarrow \cdots  |W_{N_k}\rangle \rightarrow  \cdots |W_{N}\rangle$
is calculated ($\eta _d \eta _l =1$ in the ideal case) as
\begin{eqnarray*}
p_{N}=\frac{1}{2} \prod _{k=1}^{K} \frac{2^{k}+1}{(2^{k}+2)^2} ,
\end{eqnarray*}
where $K = \log _{2} (N -2) -1$ means the number of conversions
required to breed the $|W_{N}\rangle$,
and $N_{k}=2^{k+1}+2$ satisfies $N_{k+1}=2(N_{k}-1)$.
The overhead $R_{N}=4 \times 2^{K}/p_{N}$ scales like $2^{O [(\log _2 N )^2]}$ for $N \gg 1$,
which is quasi-polynomial in the number of qubits $N$.
This is because, 
although the success probability of the conversion
$|W_{N} \rangle  |W_{N}\rangle \rightarrow |W_{2(N-1)}\rangle$
decreases as $O(1/N)$,
the size of the $W$ state grows exponentially with the number of conversions $O(\log_{2} N)$.
(The overhead will be somewhat improved by recycling.)
On the other hand,
if we generate the $|W_{N}\rangle$ by the concatenated entangling with $A^{(L)}$ as mentioned before,
the overhead $R_{N} \sim 2^{ O (N)}$ is exponential.
Furthermore,
the number of total clicks in the breeding
is $4 + 2K=3+ 2\log _{2} (N-2)$.
Thus the detector inefficiency $\eta _{d}$
and photon loss $1-\eta _l$ do not upset the scalability in the breeding scheme
though they require somewhat more resources. 
In Fig. \ref{WResource}, 
the overheads $R_{N}$ for the concatenated entangling (red $\circ$) and the breeding (blue $\times$),
respectively,
are plotted as functions of the number of qubits $N$,
where $\eta _{d} \eta _{l} = 0.5, 0.7 ,1$ from top to bottom.
As byproducts in breeding the $|W_{N}\rangle$,
the $|W_{N-2M}\rangle$ ($1 \leq M \leq N/2-1$) can also be obtained with probability 
$(N-2M)p_{N}/N$ and resources $4 \times 2^{K} [N/(N-2M)]/p_N$ by recycling.

{\it Discussion and conclusion.---}
We have considered a robust and scalable
scheme to generate large-scale entanglement webs.
We have first introduced an efficient way to generate the four-qubit $W$ state
by following the Barrett-Kok's concept,
which provides a significantly high success probability 1/2.
Then, by using the four-qubit $W$ states as seeds,
we have developed an economical breeding method
to generate an arbitrarily large $W$ state with a quasi-polynomial overhead.
The breeding method is quite simple,
and exploits an unique property of entanglement webs.
That is, a global web structure can be constructed only by a local connection.
This will provide a new perspective on multipartite entanglement.

\begin{acknowledgments}
K.F. was supported by 
JSPS Research Fellowships for Young Scientists No. 20$\cdot$2157.
\end{acknowledgments}



\begin{thebibliography}{9}
\bibitem{Cabrillo99}
C. Cabrillo, {\it et al.},
Phys. Rev. A {\bf 59}, 1025 (1999).

\bibitem{Bose99}
S. Bose, {\it et al.},
Phys. Rev. Lett. {\bf 83}, 5158 (1999).

\bibitem{Benjamin09}
S. C. Benjamin,	B. W. Lovett, and J. M.	Smith,	
Laser Photon. Rev. {\bf 3}, 556 (2009).

\bibitem{Duan10}
L.-M. Duan and C. Monroe,
Rev. Mod. Phys. {\bf 82}, 1209 (2010).

\bibitem{REO}
X.-L. Feng, {\it et al.},
Phys. Rev. Lett. {\bf 90}, 217902 (2003);
L.-M. Duan, and H. J. Kimble,
Phys. Rev. Lett. {\bf 90}, 253601 (2003);
D. E. Browne, M. B. Plenio, and S. F. Huelga,
Phys. Rev. Lett. {\bf 91}, 067901 (2003);
C. Simon, and W. T. M. Irvine,
Phys. Rev. Lett. {\bf 91}, 110405 (2003).

\bibitem{Barrett05}
S. D. Barrett and P. Kok,
Phys. Rev. A {\bf 71}, 060310(R) (2005).

\bibitem{LD0506}
Y. L. Lim, A. Beige, and L. C. Kwek,
Phys. Rev. Lett. {\bf 95}, 030505 (2005);
L.-M. Duan, {\it et al.},
Phys. Rev. A {\bf 73}, 062324 (2006).

\bibitem{Chou05}
C. W. Chou, {\it et al.},
Nature (London) {\bf 438}, 828 (2005).

\bibitem{MOM0709}
D. L. Moehring, {\it et al.},
Nature (London) {\bf 449}, 68 (2007);
S. Olmschenk, {\it et al.},
Science {\bf 323}, 486 (2009);
P. Maunz, {\it et al.},
Phys. Rev. Lett. {\bf 102}, 250502 (2009).

\bibitem{Duan05}
L.-M. Duan and R. Raussendorf,
Phys. Rev. Lett. {\bf 95}, 080503 (2005).

\bibitem{Li10}
Y. Li, {\it et al.},
Phys. Rev. Lett. {\bf 105}, 250502 (2010).

\bibitem{FT10}
K. Fujii and Y. Tokunaga,
Phys. Rev. Lett. {\bf 105}, 250503 (2010).

\bibitem{GHZ}
D. Greenberger, M. A. Horne, and A. Zeilinger,
{\it Bell's Theorem, Quantum Theory, and Conceptions of the Universe},
edited by M. Kafatos
(Kluwer, Dordrecht, 1989) p. 69.

\bibitem{Cluster}
H.-J. Briegel and R. Raussendorf,
Phys. Rev. Lett. {\bf 86}, 910 (2001).

\bibitem{Wstate}
W. D{\"u}r, G. Vidal, and J. I. Cirac,
Phys. Rev. A {\bf 62}, 062314 (2000).

\bibitem{WstateRobust}
M. Koashi, V. Bu\u{z}ek, and N. Imoto,
Phys. Rev. A 62, 050302(R) (2000);
W. D{\"u}r, Phys. Rev. A 63, 020303(R) (2001);
A. R. R. Carvalho, F. Mintert, and A. Buchleitner,
Phys. Rev. Lett. {\bf 93}, 230501 (2004).

\bibitem{WstateApp}
J. Joo, {\it et al.},
e-print arXiv:quant-ph/0204003;
J. Joo, {\it et al.},
New. J. Phys. {\bf 5}, 136 (2003);
E. D' Hondt and P. Panangaden, 
Quantum Inf. Comput. 6, 173 (2006);
S.-B. Zheng,
Phys. Rev. A {\bf 74}, 054303 (2006).

\bibitem{Miyake05}
A. Miyake, and H.-J. Briegel,
Phys. Rev. Lett. {\bf 95}, 220501 (2005).

\bibitem{WstateOpt}
T. Yamamoto, {\it et al.},
Phys. Rev. A {\bf 66}, 064301 (2002);
M. Eibl, {\it et al.},
Phys. Rev. Lett. {\bf 92}, 077901 (2004);
H. Mikami, Y. Li, K. Fukuoka, and T. Kobayashi,
Phys. Rev. Lett. {\bf 95}, 150404 (2005);
P. Walther,K. J. Resch, and A. Zeilinger,
Phys. Rev. Lett. {\bf 94}, 240501 (2005);
T. Tashima, {\it et al.},
Phys. Rev. A {\bf 77}, 030302(R) (2008); New J. Phys. {\bf 11}, 023024 (2009);
S. B. Papp, {\it et al.},
Science {\bf 324}, 764 (2009);
R. Ikuta, {\it et al.},
Phys. Rev. A {\bf 83}, 012314 (2011).


\bibitem{ExOther}
C. Yu, {\it et al.},
Phys. Rev. A {\bf 75}, 044301 (2007);
I. E. Linington, and N. V. Vitanov,
{\it ibid} {\bf 77}, 010302 (2008);
J. Song, Y. Xia, and H.-S. Song,
{\it ibid} {\bf 78}, 024302 (2008);
H.-F. Wang, {\it et al.},
J. Phys. B {\bf 42}, 175506 (2009);
D. Gon\c{t}a, and S. Fritzsche,
Phys. Rev. A {\bf 81}, 022326 (2010).

\bibitem{Dicke}
R. H. Dicke, Phys. Rev. {\bf 93}, 99 (1954).

\bibitem{Jump}
M .B. Plenio, and P. L. Knight, Rev. Mod. Phys. {\bf 70}, 101 (1998);
W. L. Power and P. L. Knight, Phys. Rev. A {\bf 53}, 1052 (1996).

\end{thebibliography}
\end{document}